\documentclass[a4paper]{article}

\usepackage{graphicx}
\usepackage{amsmath}
\usepackage{amsfonts}
\usepackage{amssymb}

\DeclareMathOperator{\Tr}{Tr}
\DeclareMathOperator{\Det}{Det}
\DeclareMathOperator{\T}{T}
\DeclareMathOperator{\IM}{Im}

\renewcommand{\d}{\partial}
\renewcommand{\ni}{\noindent}
\newcommand{\proof}{\noindent\textbf{Proof\,:\;}}

\renewcommand{\to}{\leqslant}

\title{Path integral in energy representation in quantum
mechanics}
\date{}
\author{Pavel Putrov \footnote{Saint-Petersburg State University and ITEP, Moscow. E-mail: putrov@itep.ru}}

\begin{document}

\newtheorem{prop}{Proposition}
\newtheorem{lemma}[prop]{Lemma}
\newtheorem{theorem}[prop]{Theorem}
\newtheorem{defin}{Definition}
\newtheorem{corollary}[prop]{Corollary}

\vspace{-2cm}

\maketitle

\vspace{-4cm}

\begin{flushright}
ITEP-TH-21/06
\end{flushright}

\vspace{2cm}

\begin{abstract}
In this paper we develop the alternative path-integral approach to quantum mechanics. We present a resolvent of a Hamiltonian (which is Laplace transform of a evolution operator) in a form which has a sense of ``the sum over paths'' but it is much more better defined than the usual functional integral. We investigate this representation from various directions and compare such approach to quantum mechanics with the standard ones.
\end{abstract}

\section*{Introduction}

This work can be considered as an attempt to obtain mathematically well-defined representation of the transition amplitude in quantum mechanics via the sum over paths. The representation which we obtain is actually the series each term of which contains \textit{finite} number of integrations. The space $Paths(x',x'')$ which we define and which we integrate over is much simplier than the space of trajectories $\Gamma(x',x'')=\{x(\cdot)\in C^1([0,T],V)|x(0)=x',\,x(T)=x''\}$ figurating in the usual path integral approach. We define appropriate volume form and action functional on this space and use these objects to write a ``path integral''. 

The representation which we obtain can help to realize what trajectories give the main contribution in various situations. 

The plan of the paper is the following. Section \ref{sect_main} contains main definitions and the core result. In section \ref{schr_sect} we obtain the peculiar connection between the sum over certain subset of all paths and the solution of the Schr\"odinger equation. We also give the alternative proof of our representation. In section \ref{sect_step} we consider the special type of potentials (step-like potentials) in which our representation simplifies drastically. Sections \ref{sect_semicl},\ref{sect_inst} contain considerations of some physical applications and correspondence to some usual methods in quantum mechanics. They are written in non-mathematical style. In section \ref{dim_D} we just \textit{sketch} the possible generalization(maybe not the best one) to the $D$-dimensional quantum mechanics. Finally in section \ref{conclusion} we discuss possible directions of subsequent research.

\section{The alternative representation of the transition amplitude}\label{sect_main}

We work in the 1D quantum mechanics (time is Euclidean) with the target space\footnote{By default we consider $\mathbb{R}^1$, but it is also possible to work with other 1D topology (e.g. $S^1$, segment). General things will be just the same.} $V=\mathbb{R}$. We denote coordinates on $V$ by the $x_\cdot$ variables. And we consider Hamiltonian acting on the Hilbert space $L_2(V)$ in the form $H=-\hbar^2\d^2/2+U(x)$ where the potential is a function $U\in C^1(V,\mathbb{R})$ ($U(x)\rightarrow+\infty,\;x\rightarrow\pm\infty$)and $\d\equiv \frac{d}{dx}$. The core object of our investigations is the evolution operator in the energy representation, $K_E$. In the usual (or ''time-``) representation it is $K_T=e^{-{HT}/{\hbar}}$. By energy representation we mean just a Laplace transform. That is functions of time $T$ go to functions of energy $E$. Namely 

\begin{equation}
K_E=\int\limits_0^{+\infty}dT\,K_T\,e^{{ET}/{\hbar}}=\frac{\hbar}{H-E} 
\end{equation}

\ni We shall always assume that $E$ does not coincide with any eigenvalue of $H$. When we consider the corresponding kernels we shall just specify the dependence on $x$-variables:

\begin{equation}
K_T(x'',x')=\langle x''|e^{-HT/\hbar}|x'\rangle\hspace{2 cm} K_E(x'',x')=\langle x''|\frac{\hbar}{H-E}|x'\rangle
\end{equation}

 In our approach it's more convenient to work with the function $p(x)=\sqrt{2(E-U(x))}$. The branch of the root should be chosen according to that $\IM[p(x)]>0$. Notice that we do not show explicitly the dependence of $p(x)$ on $E$. From this point let us set $\hbar=1$ (but we shall restore it in some places) for the sake of simplicity. Then we have 

\begin{equation}
K_E=-\frac{2}{\d^2+p(x)^2}
\end{equation}

\noindent Now let us introduce the operator 

\begin{equation}
\Phi=\frac{1}{\sqrt{p}}\frac{i}{p+i\d}\frac{1}{\sqrt{p}} 
\end{equation}

\noindent with the kernel

\begin{equation} \Phi(x'',x')=\frac{1}{\sqrt{p(x'')p(x')}}e^{i\int_{x'}^{x''}
p(\xi)d\xi}\cdot\theta(x''-x') \label{Phi_ker}\end{equation}

\noindent That is, it is the quasiclassical transition amplitude in the right direction. The transposed operator is the following 

\begin{equation}
\Phi^{\T}=\frac{1}{\sqrt{p}}\frac{i}{p-i\d}\frac{1}{\sqrt{p}}
\end{equation}

\begin{equation} \Phi^T(x'',x')=\frac{1}{\sqrt{p(x'')p(x')}}e^{i\int_{x''}^{x'}
p(\xi)d\xi}\cdot\theta(x'-x'') \label{Phi_T_ker}\end{equation}

And it is the quasiclassical propagator in the left direction. Notice that the kernels are convergent at large $x$ due to $\IM[p(x)]>0$.

\begin{defin}[Diagrammatic rules for operators]${}_{}$
 \newline For propagators upwards(to the right) and downwards(to the left):
\begin{center}
 \includegraphics{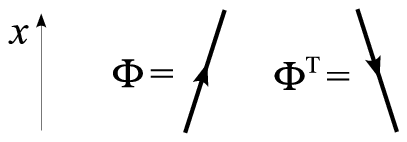}
\end{center}
 For ``reflection points'':
\begin{center}
 \includegraphics{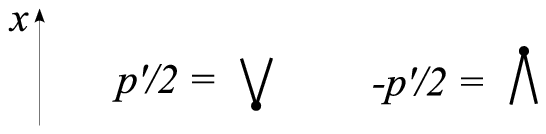}
\end{center}
where $p'\equiv \frac{\d p}{\d x}$.
\label{diagr}
\end{defin}

To write the expression corresponding to the certain diagram one should multiply operators corresponding to all elements in order determined by arrows.

\begin{lemma}

The following expansion takes place:

\begin{equation}
-iK_E=\Phi+\Phi^{\T}-\Phi^{\T}\frac{p'}{2}\Phi+\Phi\frac{p'}{2}\Phi^{\T}-\Phi^{\T}\frac{p'}{2}\Phi\frac{p'}{2}\Phi^{\T}-\Phi\frac{p'}{2}\Phi^{\T}\frac{p'}{2}\Phi+\ldots=
\label{oper1}\end{equation}

\begin{center}
 \includegraphics{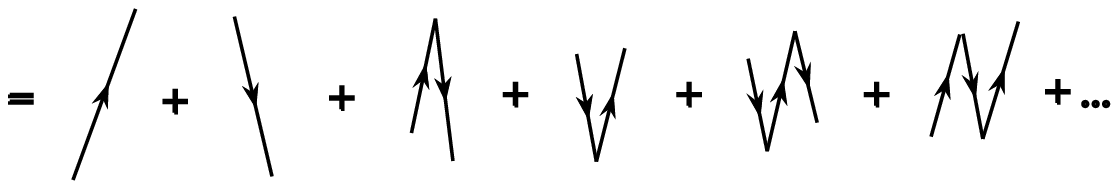}
\label{oper1_diagr}
\end{center}
\label{lemma_exp}
\end{lemma}

\proof Expansion can be written explicitly as

\[
\frac{2i}{p^2+\d^2}=\frac{1}{\sqrt{p}}\frac{i}{p+i\d}\frac{1}{\sqrt{p}}+\frac{1}{\sqrt{p}}\frac{i}{p-i\d}\frac{1}{\sqrt{p}}+\]
\begin{equation}
+\frac{1}{\sqrt{p}}\frac{i}{p+i\d}\frac{p'}{2p}\frac{i}{p-i\d}\frac{1}{\sqrt{p}}-\frac{1}{\sqrt{p}}\frac{i}{p-i\d}\frac{p'}{2p}\frac{i}{p+i\d}\frac{1}{\sqrt{p}}+\ldots
\label{oper2} \end{equation}

 Series (\ref{oper1}),(\ref{oper2}) actually can be summed explicitly, because they consist
 of the geometric series of the type
 $\frac{1}{A-B}=A^{-1}+A^{-1}BA^{-1}+\ldots$. To prove
 (\ref{oper2}) consider the operators $u_+$ and $u_-$ which are the sums of
 terms in the r.h.s. starting from $\Phi$ and $\Phi^{\T}$
 correspondingly. They satisfy the following recursive relations:
\begin{equation}
(p\pm i\d)u_{\pm}=\frac{i}{p}-i\frac{p'}{2p}(u_+-u_-)
\end{equation}
In terms of $u=u_++u_-$ and $w=u_+-u_-$ they can be rewritten as:
\begin{equation}
pu+i\d w=\frac{2i}{p}-\frac{ip'}{p}w \hspace{2cm} pw+i\d u=0
\end{equation}

\noindent then simply excluding $w$ we get the desired answer
\begin{equation}
 (p^2+\d^2)u=2i\Rightarrow u=-iK_E
\end{equation}
$\square$

 Each diagram in (\ref{oper1}) corresponds in some sense to a scheme of particle moving. And thus one can already see ''the sum over the paths``. Let us consider this more rigorously.

\begin{defin}[Paths] 
 Let $\Delta\subset V$. Then let us define $Paths^N_\Delta(x',x'')\subset \{x'\}\times\Delta^N\times\{x''\}$ (or $Paths^N_\Delta:V\times V\rightarrow 2^{\Delta^N}$)
\begin{equation}
Paths^N_\Delta(x',x'')=\{( x_0\equiv x',x_1,\cdots,x_N,x_{N+1}\equiv x'')|x_i\in\Delta,\;(x_{i+1}-x_i)(x_i-x_{i-1})<0,\;1\to i\to N\}
\end{equation} 
\begin{equation}
 Paths^N(x',x'')\equiv Paths^N_V(x',x'')\supset Paths^N_\Delta(x',x'')\;\;\forall \Delta
\end{equation} 
\begin{equation}
 Paths_\Delta(x',x'')=\bigsqcup_{N=0}^{\infty}Paths_\Delta^N(x',x'')
\end{equation} 
\end{defin}
\textit{Remark:} By default we will use the following notation for the element $P\in Paths(x',x'')$:
\begin{equation}
 P=( x_0\equiv x',x_1,\cdots,x_N,x_{N+1}\equiv x'')\in Paths^N(x',x'')\subset Paths(x',x'')
\end{equation} 
We shall always identify $ x_0\equiv x',\;x_{N+1}\equiv x''$ in the similar cases and shall not mention this. We shall call $x_k,\;1\to k\to N$ reflection points, $x'$ - the start point and $x''$ - the end point of the path $P$.

\begin{defin}[Sign function]$\rightleftarrows_{\cdot}^{(\cdot)}:Paths^N\times \{1,\ldots,N\}\rightarrow \mathbb{Z}_2$ :
\begin{equation}
 \rightleftarrows_k^{(P)}=\theta(x_{k}-x_{k-1})
\end{equation} 
\end{defin}

\begin{defin}[Action functional]
 Let us define function $S: Paths(x',x'')\rightarrow \mathbb{C}$:
\begin{equation}
S[P]=-\sum_{k=0}^{N}(-1)^{\rightleftarrows^{(P)}_{k+1}}\int\limits_{x_{k}}^{x_{k+1}}p(\xi)d\xi\equiv\int\limits_Pp(\xi)d\xi
\end{equation} 
That is the integral of $p(\xi)$ is such that each term has a positive imaginary part.
\end{defin}
\textit{Remark:} $S[P]$ is a shorten action computed on the path $P$ by the \textit{classical} formula (i.e. the formula for momentum is classical).

\begin{defin}[``Volume'' form on $\mathbf{Paths}$]
Let us first define the ``volume''\footnote{We use double quotes because this form is of course not necessary positive.} form on the $Paths^N(x',x'')$:
\begin{equation}
 \Omega_N=\bigwedge_{k=1}^Ndx_k\,\frac{p'(x_k)}{2p(x_k)}\,(-1)^{\rightleftarrows_k}
\end{equation}
Then the volume form on the $Paths(x',x'')$ can be defined as just a sum of them:
\begin{equation}
 \Omega=\sum_{N=0}^{\infty}\Omega_N
\end{equation}   
\end{defin}

Now one can easily rewrite expansion (\ref{oper1}) in the lemma \ref{lemma_exp} in terms of kernels of the operators and obtain the following:
\begin{theorem}
\begin{equation}
 \boxed{ \langle x''|\frac{1}{H-E}|x'\rangle=\frac{1}{\sqrt{p(x')p(x'')}}\int\limits_{\hbox{\tiny $Paths(x',x'')$}}\hspace{-4mm}\Omega\, e^{iS} } \label{ker1} \end{equation}
\ni or more explicitly:
\begin{equation} \boxed{ \langle x''|\frac{1}{H-E}|x'\rangle=\frac{1}{\sqrt{p(x')p(x'')}}\sum^{\infty}_{N=0}\int\limits_{\hbox{\tiny $Paths^N(x',x'')$}}\prod^{N}_{k=1}dx_k\frac{p'(x_k)}{2p(x_k)}(-1)^{\rightleftarrows_k}\,e^{\hspace{2mm}i\int\limits_{P}p(\xi)d\xi} } \label{ker} \end{equation}
\end{theorem}

\noindent \textit{Remarks:} Sum over $N$ just corresponds the sum in (\ref{oper1}).  The $N$-th term in the sum (\ref{ker}) contains two terms from the sum (\ref{oper1}) with $N$ being the number of $p'$ entering into them.  Convolution over the points $\{x_k\}_{k=1}^{N}$ is performed over the set ${\{(x_{i+1}-x_i)(x_i-x_{i-1})<0\}}$. These inequalities are produced by the $\theta$-functions from the kernels (\ref{Phi_ker}) and (\ref{Phi_T_ker}).  The exponential in (\ref{ker}) is just the product of the exponentials entering into the kernels of $\Phi$ and $\Phi^T$. 

This goes along with our diagrammatic rules from def. \ref{diagr}. 
\begin{defin}[Diagrammatic rules for computing (\ref{ker})]

Propagators:
\begin{center}
 \includegraphics{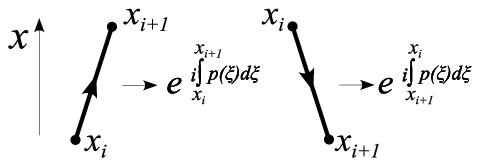}
\end{center}
Reflection points:
\begin{center}
 \includegraphics{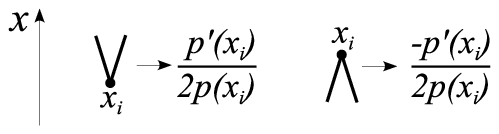}
\end{center}
Start and end points:
\begin{center}
 \includegraphics{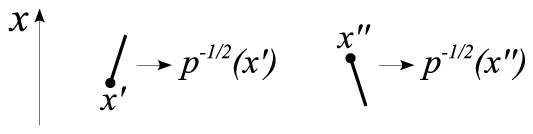}
\end{center}
And one should integrate over the positions of reflection points in such a way that the structure of arrows does not change (i.e. the number of arrows and their orientation remain the same)
\end{defin}
Then one can write (\ref{ker}) as
\begin{center}
 \parbox{3cm}{\vspace{-25mm}$\langle x''|\frac{1}{H-E}|x'\rangle=$}\vspace{15mm}\includegraphics{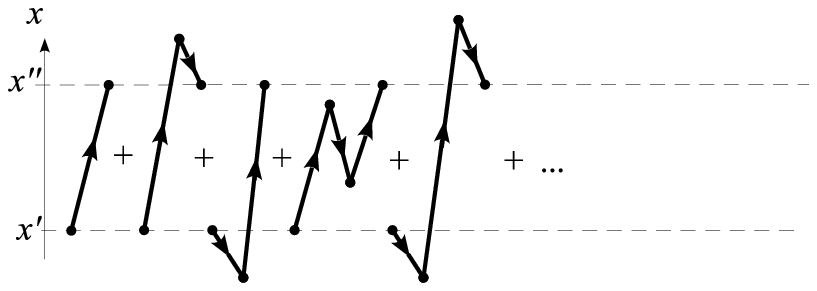}
\end{center}

\begin{corollary}
 There is a following representation for the partition function:
\begin{equation}
 Z_E=\Tr\frac{1}{H-E}=\int\limits_V\frac{dx}{p(x)}\int\limits_{\hbox{\tiny $Paths(x,x)$}}\hspace{-4mm}\Omega\, e^{iS}  \label{Z1} \end{equation}
\end{corollary}

\section{Connection with the Schr\"{o}dinger equation}\label{schr_sect}

Let us consider the function 
\begin{equation}
 I_R(x)=\hspace{-4mm}\int\limits_{Paths_{[x,+\infty)}(x,x)}\hspace{-4mm}\Omega\,e^{iS}
\end{equation} 
i.e. the sum over the paths with reflection points lying entirely to the right side of $x$:
\begin{center}
 \includegraphics{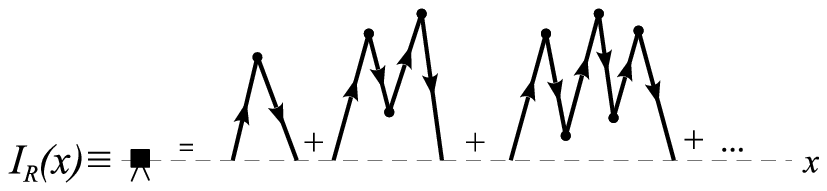}
\end{center}
One can obtain the differential equation on the $I_R(x)$ considering the small shift of $x$:
\begin{center}
 \includegraphics{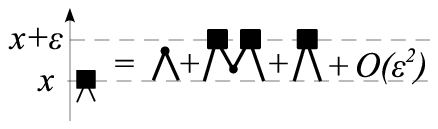}
\end{center}
\begin{equation}
\frac{dI_R}{dx}=\frac{p'}{2p}(1-I_R^2)-i\frac{2p}{\hbar}I_R
\label{formula_eq_r}
\end{equation}
And analogously for the function
\begin{equation}
 I_L(x)=\int\limits_{Paths_{(-\infty,x]}(x,x)}\Omega\,e^{iS}
\end{equation}
we have
\begin{equation}
\frac{dI_L}{dx}=\frac{p'}{2p}(1-I_L^2)+i\frac{2p}{\hbar}I_L
\label{formula_eq_l}
\end{equation}
One should add an obvious boundary conditions to these equations\footnote{For the case when $V=[a,b]$ they should be changed to $I_L|_{a}=-1,\;I_R|_{b}=-1$}: $I_L|_{-\infty}=0,\;I_R|_{+\infty}=0$. $E$ plays the role of a parameter.

One can obtain the following result just by making several changes of variables:
\begin{prop}
Equations (\ref{formula_eq_l}) and (\ref{formula_eq_r}) are actually equivalent to the Schr\"{o}dinger one. That is

\begin{equation}
I_R=-\frac{(\log\psi_R)'-\frac{ip}{\hbar}}{(\log\psi_R)'+\frac{ip}{\hbar}}\hspace{2cm}
I_L=-\frac{(\log\psi_L)'+\frac{ip}{\hbar}}{(\log\psi_L)'-\frac{ip}{\hbar}}
\label{relation} \end{equation}

\noindent where $\psi_L,\psi_R$ are the solutions to the Schr\"{o}dinger equation

\begin{equation} \psi''=-\frac{p^2}{\hbar^2}\psi,\;\;p(x)^2=2(E-U(x))
\label{schr}\end{equation}

\noindent with the boundary conditions\footnote{When $V=[a,b]$ they should be modified to $\psi_L|_{a}=0,\;\psi_R|_{b}=0$.}
$\psi_L|_{-\infty}=0,\;\psi_R|_{+\infty}=0$.
\end{prop}

With the help of $I_L$ and $I_R$ one can easily write the sum over paths for the partition function. That is
\begin{equation}
\int\limits_{\hbox{\tiny $Paths(x,x)$}}\hspace{-4mm}\Omega\, e^{iS} =\frac{1+I_L(x)+I_R(x)+I_L(x)I_R(x)}{1-I_L(x)I_R(x)}
\label{Z_IR_IL}
\end{equation} 
One should just perform the summation(of geometric series) in $x$ to obtain this.

Let us introduce the Wronskian for the pair ($\psi_L$,$\psi_R$): $W(E)=\left(\psi_L'(x)\psi_R(x)-\psi_R'(x)\psi_L(x)\right)$. Wronskian has zeroes in the eigenvalues of $H$. Then substituting (\ref{relation}) into (\ref{Z_IR_IL}) and using that $W'(E)\equiv\frac{dW(E)}{dE}=2\int_V\psi_L(x)\psi_R(x)dx$ one can get:
\begin{equation}
\int\limits_V\frac{dx}{p(x)}\int\limits_{\hbox{\tiny $Paths(x,x)$}}\hspace{-4mm}\Omega\, e^{iS}  =-\frac{W'(E)}{W(E)}=\Tr\frac{1}{H-E}=Z_E
\end{equation} 
\ni This computation can be considered as the alternative proof of (\ref{Z1}).

As a consequence if we know the solutions of the Schr\"odinger equation  then we know $I_R$ and $I_L$ which contain information about trajectories contributing to (\ref{Z1}). Derivatives of $I_R$ and $I_L$ give us information about trajectories reflecting in the vicinity of $x$.

The analogous computations (they can be found in the app. \ref{schr_proof}) can be made for the transition
amplitude. In this case we should introduce function
$J_R(x,x'')$(or $J_L(x,x')$) for the sum of all trajectories
connecting $x$ and $x''$ and lying to the right of $x$:
\begin{equation}
 J_R(x,x'')=\hspace{-4mm}\int\limits_{Paths_{[x,+\infty)}(x,x'')}\hspace{-4mm}\Omega\,e^{iS}
\end{equation} 
 It satisfies the following equation: 
\begin{equation}
J_R'(x,x'')=-i\frac{p}{\hbar}J_R(x,x'')-\frac{p'}{2p}J_R(x,x'')I_R(x),\;\;\;x<x'' \label{formula_eq_J}
\end{equation}
with the boundary condition
\begin{equation}
 J_R(x'',x'')=I_R(x'')+1 \label{formula_eq_J_2}
\end{equation} 

Then using this function one can obtain:

\begin{equation}
\frac{1}{\sqrt{p(x')p(x'')}}\hspace{-4mm}\int\limits_{Paths(x,x'')}\hspace{-4mm}\Omega\,e^{iS[P]}=-2\frac{\psi_L(x')\psi_R(x'')}{W(E)}=\sum_n\frac{\psi_n(x')\psi_n(x'')}{E_n-E}=\langle
x'|\frac{1}{H-E}|x''\rangle=K_E(x',x'')\label{formula_schr_K} \end{equation}

\noindent for $x'<x''$. This is an alternative proof of (\ref{ker1}), (\ref{ker}).

Let us notice also some consequence following from what we have
just got. If one wants to calculate path integral over the
trajectories with $x$ not going out from some fixed segment, one
can construct the corresponding kernel from solutions of the
Schr\"{o}dinger equation with boundary conditions $\d\log\psi/\d
x=\pm ip/\hbar$ (such that $I_{L,R}=0$). This differs from the
standard case with infinitely high potential walls at the
boundary, for which we should imply $\psi=0$ at the boundary
($I_{L,R}=-1$).

\section{Step-like potentials}\label{sect_step}
One can easily notice that due to the factors $\frac{p'}{2p}$ the integrals in (\ref{ker}) are simplified drastically in the domains of constant potential. Then one can consider step-like potential $U(x)$ with $U(x)=U_j,
x\in\Delta_j\subset V\;(\Delta_j=[y_{j-1},y_j])$ (i.e. $U(x)$ has discontinuities in points $y_j\;(y_{j}>y_{j+1}),\;j\in \mathbb{Z}$). We shall denote the length of
the interval $\Delta_j$ by $a_j$ and $p_j=\sqrt{2(E-U_j)}$.

What we need is to find the integral in the vicinity of some point $y_j$ where potential changes from $U_j$ to $U_{j+1}$. Let us regularize the step. That is, assume that in the interval $\Sigma_\epsilon=[y_j-\epsilon,y_j+\epsilon]$ we have some smooth function $p(x)=\sqrt{2(E-U(x))}$ such that $p(y_j-\epsilon)=p_j,\;p(y_j+\epsilon)=p_{j+1}$. Let us consider first the integral corresponding to the case when path somehow reflects from this step. Then we are interested in the quantity:
\begin{equation}
 -\alpha_j=\lim_{\epsilon\rightarrow 0}\hspace{-5mm}\int\limits_{Paths_{\Sigma_\epsilon}(y_j-\epsilon,y_j-\epsilon)}\hspace{-5mm}\Omega\,e^{\frac{iS}{\hbar}}=\hspace{-5mm}\int\limits_{Paths_{\Sigma_\epsilon}(y_j-\epsilon,y_j-\epsilon)}\hspace{-5mm}\Omega\label{step_alpha_int}
\end{equation} 
Because the last expression obviously does not depend on $\epsilon$. Note that this also can be considered as $\hbar\rightarrow\infty$ limit (because what we use is that $e^{iS/\hbar}\rightarrow 1$), that is the step is infinitely quantum system (in contradistinction to the domains of constant potential which can be considered as classical), only volume form $\Omega$ on the $Paths$ survives. Computations in the app.\ref{tanh_proof} give that
\begin{equation}
 \alpha_j=\frac{p_{j+1}-p_j}{p_{j+1}+p_j}\label{step_alpha}
\end{equation} 
Analogously one can compute that
\begin{equation}
 \lim_{\epsilon\rightarrow 0}\hspace{-5mm}\int\limits_{Paths_{\Sigma_\epsilon}(y_j+\epsilon,y_j+\epsilon)}\hspace{-5mm}\Omega\,e^{\frac{iS}{\hbar}}=+\alpha_j
\end{equation}  
\begin{equation}
 \lim_{\epsilon\rightarrow 0}\hspace{-5mm}\int\limits_{Paths_{\Sigma_\epsilon}(y_j-\epsilon,y_j+\epsilon)}\hspace{-5mm}\Omega\,e^{\frac{iS}{\hbar}}=1-\alpha_j\hspace{15mm}\lim_{\epsilon\rightarrow 0}\hspace{-5mm}\int\limits_{Paths_{\Sigma_\epsilon}(y_j+\epsilon,y_j-\epsilon)}\hspace{-5mm}\Omega\,e^{\frac{iS}{\hbar}}=1+\alpha_j
\end{equation}  
Notice that these quantities coincide with well-known reflection/absorption coefficients of the step.

\begin{defin}[Paths for step-like potential] 
\[
\widetilde{Paths}(x',x'')=\left\{\left.(j_1,\ldots,j_N)\right|x'\in\Delta_{j'}\Rightarrow j_1\in\{j'-1,j'\},\right.
\]
\begin{equation}
\left.j_{k+1}\in\{j_k-1,j_k+1\}\,(1\to k \to N),x''\in\Delta_{j''}\Rightarrow j_N\in\{j''-1,j''\}\right\}
\end{equation}
\end{defin}
\textit{Remarks:} By $N(P)$ we shall denote the $N$ for element $P\in \widetilde{Paths}$ figuring in this definition. We shall sometimes consider indicies $j_0\in\{j'-1,j'\}\setminus\{j_1\}$ and $j_{N+1}\in\{j''-1,j''\}\setminus\{j_{N+1}\}$. Elements of $\widetilde{Paths}$ can be thought of as random-walks between the points $y_j$.

\begin{defin}[reflection/absorption coefficients] Let $P=(j_1,\ldots,j_N)\in \widetilde{Paths}(x',x'')$. 
\begin{equation}
 A_k[P]=\left\{\begin{array}{rcl}
 1-\alpha_{j_k} &\;,\;&j_{k-1}=j_k-1,\,j_{k+1}=j_k+1\\
1+\alpha_{j_k} &\;,\;&j_{k-1}=j_k+1,\,j_{k+1}=j_k-1\\
-\alpha_{j_k} &\;,\;&j_{k-1}=j_k-1,\,j_{k+1}=j_k-1\\
\alpha_{j_k} &\;,\;&j_{k-1}=j_k+1,\,j_{k+1}=j_k+1\\
\end{array}
 \right.
\end{equation} 
\ni for $1\to k\to N$
\end{defin}
\begin{defin}[Action] Let $P=(j_1,\ldots,j_N)\in \widetilde{Paths}(x',x'')$ and $x'\in\Delta_{j'},\;x''\in\Delta_{j''},\;[y_{j_k},y_{j_{k+1}}]=\Delta_{l_k}$.
\begin{equation}
 S[P]=p_{j'}|x'-y_{j_1}|+\sum_{k=1}^{N-1}p_{l_k}a_{l_k}+p_{j''}|y_{j_N}-x''|
\end{equation} 
\end{defin}

Taking into account everything above in the text one can deduce the following theorem for the case of step-like $U(x)$:
\begin{theorem}
\begin{equation}
 \langle x''|\frac{1}{H-E}|x'\rangle=\frac{1}{\sqrt{p(x')p(x'')}}\sum_{P\in \widetilde{Paths}(x',x'')}\prod_{k=1}^{N(P)}A_k[P]\,e^{iS[P]} \label{ker_step} 
\end{equation} 
\end{theorem}

Note that one can represent the arbitrary potential as a limit of step-like ones and obtain (\ref{ker}) from (\ref{ker_step}).

One can consider the simplest case when we have just one interval of constant potential bounded by infinitely high walls (i.e. particle in the box). Then our representation has a sense of Poisson resummation of the result obtained via the Schr\"odinger equation. Thus the general case can be considered as generalization of the Poisson resummation formula.

\section{Semiclassical limit}\label{sect_semicl}

We can restore the transition amplitude $K_T(x',x'')$ (for which the usual path integral representation is written) in the time-representation in the following simple way (this is just the inverse Laplace transform):
\begin{equation}
 K_T(x',x'')=-\hspace{-5mm}\oint\limits_{\hbox{\parbox{15mm}{\vspace{-3mm}\center\tiny eigenvalues\\of $H$}}}\hspace{-5mm}\frac{dE}{2\pi i\hbar}\,K_E(x',x'')\,e^{-ET/\hbar}=\nonumber
\end{equation} 
\begin{equation}
 =-\oint\frac{dE}{2\pi i\hbar}\frac{1}{\sqrt{p(x')p(x'')}}\sum^{\infty}_{N=0}\int\limits_{\hbox{\tiny $Paths(x',x'')$}}\prod^{N}_{k=1}dx_k\frac{p'(x_k)}{2p(x_k)}(-1)^{\rightleftarrows_k}\,e^{\hspace{2mm}i\int\limits_{P}p(\xi)d\xi} 
\label{formula_K}
\end{equation} 

 Let us consider now the semiclassical limit of this formula for the transition amplitude $K_T(x',x'')$. For each trajectory $P\in Paths(x',x'')$ we can take the integral of the corresponding term over $E$ by the saddle-point method asymptotically with $\hbar\rightarrow 0$:

\[ -\oint
\frac{dE}{2\pi i\hbar}\frac{1}{\sqrt{p(x',E)p(x'',E)}}A[P](E)e^{\frac{i}{\hbar}S[P](E)}e^{-ET/\hbar}
=\] \begin{equation} =\sqrt{\frac{1}{2\pi\hbar (\partial
T_P/\partial E)}}\frac{A[P]}{{\sqrt{p(x')p(x'')}}}
e^{-\frac{1}{\hbar}(-iS[P](E)+ET)}\left(1+c_1\hbar+c_2\hbar^2+\ldots\right)|_{i\frac
{\partial }{\partial E}S[P](E)\equiv T[P](E)=T}
 \label{asympt} \end{equation}

Where $A[P]=\prod_{k=1}^N(-1)^{\leftrightarrows_k}\frac{p'(x_k)}{2p(x_k)}$. Let us denote $E_{(Eucl)}\equiv -E,\;U_{(Eucl)}(x)\equiv -U(x),\;p_{(Eucl)}(x)\equiv\sqrt{2(E_{(Eucl)}-U_{(Eucl)}(x))}=\sqrt{2(U(x)-E)}$.  One can show that the saddle point is always such that $E$ is real and $E_{(Eucl)}>\max\limits_{x\in P}U_{(Eucl)}(x)\Rightarrow p_{(Eucl)}(x)>0$ on the $P$. Hence in the exponential we obtain the full Euclidean action\footnote{Going from the shorten action to the full and back is actually the Legendre transform: $S^{sh}(E)-ET=S^{full}(T),\;T=\partial S^{sh}/\partial E,\;E=-\partial S^{full}/\partial T$.} $S_{(Eucl)}^{full}=-iS[P](E)+ET=\int_P p_{(Eucl)}(\xi)d\xi-E_{(Eucl)}T$. 

The equation for the saddle point is $T=i\frac
{\partial }{\partial E}S[P](E)=\int_P\frac{d\xi}{p_{Eucl}(\xi)}$. The r.h.s. is the classical formula for computing the time on given trajectory. That is this equation says that energy $E$ is such that the time of moving along the path $P$ is $T$. One can think about such movement as ''locally-classical``, because it is classcal everywhere except possible non-classical reflection points $x_k$, where the potential $U_{(Eucl)}(x)$ is smaller than $E_{(Eucl)}$. From
the point of view of the functional integral, instead of the ordinary expansion near the pure classical trajectories we expand near the ''locally-classical`` ones which set is more rich\footnote{Of course after integration over reflection points $x_k$ in the limit $\hbar\rightarrow 0$ in the continuous potential only the pure classical trajectories survive, as written below.}. The infinite series in $\hbar$ in (\ref{asympt}) correspond to the fluctuations around these trajectories. Every term can be written in terms of derivatives of $S[P](E)$ and $A[P](E)$ w.r.t. $E$.

Consider the transition amplitude in the case when pure classical trajectory have no turning points. Then in series
$(\ref{formula_K})$ the term with $N$ reflection points contributes only to the $\mathcal{O}(\hbar^{[(N+1)/2]})$ part of the asymptotic expansion of $K_T(x',x'')$ in the limit $\hbar\rightarrow 0$. These terms are contributed by paths in which the reflection points are situated in small vicinity ($\sim\hbar$) of the classical path.

The leading order of the expansion of $N=0$ term gives the right one-loop answer. It follows from that
\begin{equation} p(x'')p(x')\frac{\partial T[P]}{\partial
E}=\Det_{t_1,t_2}\frac{\delta^2S^{full}[x]}{\delta x(t_1) \delta x(t_2)}
\end{equation} 
\noindent To prove this, one should represent the determinant as a solution to differential equation in a standard way. In the case when the pure classical trajectory has the turning points, this statement should be modified slightly (the reason can be found below), roughly speaking, corrections are given by the reflections which does not lie in the very small vicinity of the turning point.

 One knows that there is no pure classical instanton multi-kink solution in the double-well system. But it turns
out that such configurations are close to the obtained ''locally-classical`` trajectories. The arguments can be found in the next section.

 Instead of taking the integral over $E$ as before, we can first perform the integrations over reflection points $x_k$
semiclassically. First, we should deform the contour of integration over $E$ so that it will go closely to the real axis (where the poles lie which we are interested in): $E=\mathcal{E}\pm i\epsilon$, where $\mathcal{E}$ is real. The
main contribution will be the one from the configurations where all $x_k$ lie in the vicinities of the classical turning points $\mathcal{E}=U(x)$. We can sum the reflections in one turning point\footnote{Practically it is investigating the corresponding equations in the vicinities of turning points.}, and obtain the semiclassical summation rules for $K_E$ or $Z_E$: we should sum over all trajectories with reflections only in roots \footnote{To obtain the expression analytically correct in $E$ we should consider also complex roots.} of $\mathcal{E}=U(x)$ with factors
$+i$ or $-i$ for reflections when trajectory goes from region with higher or lower then $\mathcal{E}$ potential correspondingly. In the case of the single-well (two roots) this will give us the expression of the type $\frac{(\cdots)}{1+e^{\frac{i}{\hbar}\oint pdx}}$ whose pole structure gives us just the Bohr-Sommerfeld quantization (By $(\cdots)$ here and below we denote something analytic in $E$).

\section{Instantons}\label{sect_inst}
Now we shall consider the case of the double-well potential. Let the energy be such that $\mathcal{E}=U(x)$ has 4 roots (we shall denote them $x_1,x_2,x_3,x_4$ from the left to the right). Let $S=\int_{x_1}^{x_2}pdx=\int_{x_3}^{x_4}pdx,\;R=\int_{x_2}^{x_3}|p|dx$. Then the described summation gives us the expression 
\begin{equation}
 Z_E\sim\frac{(...)}{e^{-2R/\hbar}+(e^{2iS/\hbar}+1)^2}\label{pole_shift}
\end{equation} 
For oscillator $S=2\pi i \frac{E}{\hbar\omega}$. Let $E=E^0+\Delta E,\;E^0=\hbar\omega(n+1/2),\Delta E\ll E$. Then the obtained equation gives $\left(\Delta E \frac{2\pi}{\hbar\omega}\right)^2=e^{-2R/\hbar}\Rightarrow\Delta E=\pm\frac{\hbar\omega}{2\pi}e^{-R/\hbar}$, which is exactly the known answer. The term $e^{-2R/\hbar}$ in the denominator which shifts the pole comes from the geometric series, the $N$-th term of which contains $e^{-2NR/\hbar}$. It corresponds to the path going $2N$ times from one top to the another and so containing $2N$ instantons.

There is the well-known computation of the energy splitting (e.g.\cite{coleman}) which assumes that the main contribution in the path integral for large $T$ is given by $N$-kink solutions with arbitrary positions of the kinks. Let us consider the classical trajectories in the upside-down potential. The classical trajectory which connects one top with the another contains only one instanton for any finite $T$. If we consider the trajectories with energy lower than the maximum of the potential then there are possible solutions with $N$ instanton, but the solution is
periodic and, hence, the time intervals between all of them are the same.

Let us show how it can be clarified that for large $T$ the main contribution is indeed given by the trajectories where particle stays an arbitrary time of the top of the potential. When $\hbar\rightarrow 0$ in (\ref{formula_K}), the main contribution is given by the trajectories with the turning points $E=U(x)$. As it was already mentioned above, every such turn is given actually by the sum over arbitrary combinations of the reflections in the small vicinity of the root. We can estimate the extra time which the particle spends in the vicinity of this point (in assumption that it is small w.r.t. $T$): 
$\Delta T\sim \left[\hbar m/(U')^2\right]^{1/3}\propto\hbar^{\frac{1}{3}}e^{\frac{\omega T}{3N}}$ for $N$ instantons. When $\hbar\rightarrow 0$ (and $T$ is fixed), of course, $\Delta T\rightarrow 0$. But if $\hbar$ even is very small but fixed \footnote{Formally it is enough that $\hbar\rightarrow 0$ as power of $T$.}, the ratio $\frac{\Delta T}{T}$ grows very fast when $T\rightarrow\infty$. Then, the obtained estimation becomes wrong and $\frac{\Delta T}{T}\rightarrow 1$ (goes to saturation). Thus quantum reflections in the vicinity of the top are the very thing which makes particle to stay an arbitrary time at the top of the potential when the energy is close to zero (time goes to infinity).

In the appendix \ref{inst_comp}, we show how the standard computation of the energy splitting trough instantons can be reproduced using the obtained representation for the transition amplitude. Simply speaking, geometric series shifting the pole in the energy representation turns to exponential series in the time representation.

Note that one of the main features of the Laplace transform is that the contraction w.r.t. time turns to the multiplication. Thus the functional integral as a sum of different sequences of events during the fixed time turns to the sum of sequences of independent events, i.e. so that particle at each point of its path ''does not remember`` the prehistory, and the weights corresponding to the paths in sum have multiplicative structure (up to factors at the
ends): $A{[P_1\sqcup P_2]}e^{i{S{[P_1\sqcup P_2]}}}=A{[P_1]}e^{i{S{[P_1]}}}\cdot A{[P_2]}e^{iS{[P_2]}}$. Such a form also provides rather easy description of the system which consists of some more basic subsystems. In the case of the
double-well potential, these are two quadratic wells and the intermediate region. In the case of the step-like potential, these are the steps themselves.

\section{Straightforward generalization to $D$-dimensional quantum mechanics}\label{dim_D}

Representation (\ref{oper2}) of formula (\ref{ker}) hints one to a possible generalization to the $D$-dimensional quantum mechanics. To obtain (\ref{oper2}) in $D$ dimensions as it is, we should make an extension to the Clifford algebra. Namely, nothing will change in the proof of $(\ref{oper2})$ if we make the following substitution:

\begin{equation}
\d\rightarrow \hat{\d}\equiv \gamma_{a}\d_a,\;\;\;\;p'\rightarrow (\hat{\d}p),\;\;\;\;\{\gamma_a,\gamma_b\}=2\delta_{ab},\;\;\;\;a,b=1\ldots D
\end{equation}

There is nothing bad in making such an extension because the resulting answer will be in the $\mathbb{C}$-component. In matrix representation we can take a $2^{-[D/2]}\Tr_{\gamma}$ of each term to write the formula just in $\mathbb{C}$:

\begin{equation}
\frac{2i}{p(x)^2+\Delta}=2^{-[D/2]}\Tr_{\gamma}\left[\frac{1}{\sqrt{p}}\frac{i}{p+i\hat{\d}}\frac{1}{\sqrt{p}}+\frac{1}{\sqrt{p}}\frac{i}{p-i\hat{\d}}\frac{1}{\sqrt{p}}+
\frac{1}{\sqrt{p}}\frac{i}{p+i\hat{\d}}\frac{(\hat{\d}p)}{2p}\frac{i}{p-i\hat{\d}}\frac{1}{\sqrt{p}}+\ldots
\right] \label{oper_g}
\end{equation}

It is possible to represent the kernels of $i/(p\pm i\hat{\d})$ in the following way.

One knows the path integral representation for the Dirac propagator (\cite{P1,P2}):
\begin{equation}
G_m(x'',x')\equiv\langle
x''|\frac{1}{m+\hat{\d}}|x'\rangle=\sum_{P\in \{\hbox{\tiny{paths}
$x'\rightarrow
x''$}\}}e^{-mL_P}\cdot\mathcal{P}\exp\left\{\int_Pds\frac{1}{4}\omega_{ab}(s)[\gamma_{a},\gamma_{b}]\right\}
\label{polyakovs}
\end{equation}

Here $L_P=\int_Pds$ is the length of $P$, $s$ is the natural parameter on the curve $P$ and $\omega_{ab}=\dot{x}_{[a}\ddot{x}_{b]}$ is the angle speed of the unit tangent vector. The second factor in (\ref{polyakovs}) is the so-called spin factor. We can generalize it for our purposes to:

\begin{equation}
G_{[p]}(x'',x')\equiv\langle
x''|\frac{i}{p+i\hat{\d}}|x'\rangle=\sum_{P\in
\{\hbox{\tiny{paths} $x'\rightarrow
x''$}\}}e^{i\int_Pdsp(x(s))}\cdot\mathcal{P}\exp\left\{\int_Pds\frac{1}{4}\omega_{ab}(s)[\gamma_{a},\gamma_{b}]\right\}
\label{pol_mod}
\end{equation}

This can be proved using (\ref{polyakovs}) by varying $p(x)\rightarrow p(x)+\epsilon\phi(x)$ in the r.h.s. of
(\ref{pol_mod}):

\begin{equation}
\int dy\phi(y)\frac{\delta}{\delta p(y)}G_{[p]}=iG_{[p]}\phi
G_{[p]}\;\;\Leftrightarrow\;\;\frac{\delta}{\delta
p(y)}G^{-1}_{[p]}=-i\delta(x-y)\;\;\Leftrightarrow\;\;
G_{[p]}=\frac{1}{\textrm{Const}-ip(x)}
\end{equation}

\noindent and then if (\ref{pol_mod}) is true for $p(x)=im$ (it becomes (\ref{polyakovs})) it is true for arbitrary $p(x)$.

Now we can use it in (\ref{oper_g}) - our generalization of (\ref{oper1}) - to write the generalization of (\ref{ker}) for $D>1$: \tiny
\[\langle x'|\frac{\hbar}{H-E}|x''\rangle=\]
\begin{equation} 2^{-[D/2]}\oint\!\frac{1}{\sqrt{p(x')p(x'')}}\sum^{\infty}_{N=0}\sum_{r=0}^{1}(-1)^{[\frac{N+r}{2}]}\!\int\!\!\prod^{N}_{k=1}\!dx_k\,\Tr_{\gamma}\!\mathcal{P}\frac{(\hat{\d}p)(x_k)}{2p(x_k)}\prod_{j=0}^N\!
\sum_{\parbox{5 mm}{$P_j\in$\\ $\{\textrm{paths}$
\\ \mbox{$x_j\rightarrow x_{j+1}\}$}}}\!
\exp\left\{\frac{1}{\hbar}S[P_j]+\frac{(-1)^{r+j}}{4}\!\int\limits_{P_j}\!\!ds\omega_{ab}(s)[\gamma_{a},\gamma_{b}]\right\}
\label{formula_K_D} \end{equation}
\normalsize

\noindent where $S[P]=\left(i\int_Pp(x(s))ds-ET\right)$ and we have made an identification $x_0\equiv x',\;x_{N+1}\equiv x''$ again.

Thus we obtained the representation of the transition amplitude as the sum over trajectories in the $x$-space (they are indeed geometrical trajectories, not functions $x(t)$ as they are in the standard approach). Again, the weights for trajectories have multiplicative structure: $A{[P_1\sqcup P_2]}e^{i{S{[P_1\sqcup P_2]}}}=A{[P_1]}e^{i{S{[P_1]}}}\cdot A{[P_2]}e^{iS{[P_2]}}$. The following consequences (independence of the prehistory, etc.) were already discussed in the section $6$.

The path integrals in (\ref{polyakovs}) and (\ref{pol_mod}) are not well-defined, but they have the following feature: the paths that contribute to the integral have the Hausdorff dimension $1$ (contrary to the path integral for the bosonic propagator where paths have dimension 2). This is due to the spin factor which suppresses the trajectories with large curvature, and so we can count only smooth trajectories. In $D=1$ case\footnote{It can not
be obtained straightforwardly from the general formulae, they are written for $D>1$.} considered before the trajectories can not even change the direction.

We can realize both (\ref{formula_K_D}) and (\ref{ker}) as the expansion over $N$ - the number of points of discontinuity of the tangent to a path as the curve in $x$-space.

\section{Conclusion}\label{conclusion}
Here we note several more or less interesting (from the author's subjective point of view) ways of the subsequent research: 

1)Try to find the generalization (better than in sect. $\ref{dim_D}$) to the $D$-dimensional quantum mechanics (i.e. increase target-space dimension). 

2)Try to generalize to QFT (i.e. increase world-volume dimension). 

Negative results for both cases also can be interesting. One can try to formulate ``NO-GO-to-the-higher-dimensions'' theorems. 

3) Find explicit connection of the representation (\ref{ker1}) with usual functional integral. Let us define the space of functions $\Gamma(x',x'')=\{(T>0,x(\cdot)\in C^1([0,T],V))|x(0)=x',\,x(T)=x''\}$ (the time of moving is not fixed). Then we can define the projection map $\pi:\Gamma(x',x'')\rightarrow Paths(x',x'')$ - one should forgot about the dependence of $x(t)$ on $t$, and leave only the $x$-positions of local maxima and minima\footnote{In domains of constant potential one should forget them too.} - $x_k$. Then it will be interesting to obtain something like the following (maybe with some corrections):
\begin{equation}
 \frac{1}{\sqrt{p(x')p(x'')}}\int\limits_{\Lambda}\Omega\, e^{iS}=\int\limits_{\pi^{-1}\Lambda} dT\mathcal{D}xe^{-S^{full}[x]+ET},\hspace{2cm}\forall\Lambda\subset Paths(x',x'')
\end{equation} 
i.e. measure $\mu:\mu(\Lambda)=\int_{\Lambda}\Omega e^{iS}$ that we construct on the $Paths(x',x'')$ can be thought of as pushforward $\mu=\pi_*\tilde{\mu}$ of the hypothetical measure $\tilde{\mu}$ on the $\Gamma(x',x'')$ under projection $\Gamma(x',x'')\stackrel{\pi}{\rightarrow}Paths(x',x'')$. The motivation for this equality is the locality of functionals on the trajectories at the both sides of equation. When we perturb the potential $U(x)$ in the vicinity of some point, then only the part of the sum containing the trajectories passing through this point will change. In this paper we have proved this equality for $\Lambda=Paths(x',x'')$. 

4)Try to find some connections with theories of random walks.

\section*{Acknowledgements}
 Author would like to thank Andrei Mironov, Alexei Morozov and
 Andrei Losev for helpful questions and discussion.

 The work was supported in part by the Grant of
Support for the Scientific Schools 8065.2006.2 and
 RFFI grant 04-02-16538.

\appendix

\vspace{1cm} {\LARGE \bf Appendix}

\section{Correspondence with common computation of energy splitting trough instantons} \label{inst_comp}

Consider the transition amplitude in the energy representation in
the double-well potential (which is upside-down in the Euclidean
case) going from one top to the other one. Let tops be situated at
points $-a$ and $a$. We will be interested in the case when
$T\rightarrow\infty$ and in the most left pole in $E$. Accordingly
to what was said before in $(\ref{formula_K})$, we neglect the
trajectories with reflections outside the intervals $[-a-b,-a+b]$
and $[a-b,a+b]$. Let us introduce the basic components of our sum.

1) The ''instanton operator`` with the kernel:

\begin{equation}  I(x'',x')=
\exp\left(\frac{i}{\hbar}(-1)^{\theta(x'-x'')}\int_{x'}^{x''}p(\xi)d\xi\right)\end{equation}

\noindent where $x_1$ and $x_2$ are lying in different segments
$[-a-b,-a+b]$ and $[a-b,a+b]$. These two vicinities are actually
symmetric, and we can consider for simplicity just one copy (let
us call it $\Delta$) and remember only that the total number of
instantons is odd.

2) The sum of trajectories lying in $\Delta$ between two
instantons:

\begin{equation}
Y(x'',x')=\hspace{-4mm}\int\limits_{Paths_\Delta(x',x'')}\hspace{-4mm}\Omega\,e^{iS}
\end{equation}

3) The start and end parts:

\begin{equation}
Y_{in}(x'')=\hspace{-4mm}\int\limits_{Paths_\Delta(-a,x'')}\hspace{-4mm}\Omega\,e^{iS}
\end{equation}

\begin{equation}
Y_{out}(x')=\hspace{-4mm}\int\limits_{Paths_\Delta(x',a)}\hspace{-4mm}\Omega\,e^{iS}
\end{equation}

\begin{center}
  \includegraphics[bb=5mm 235mm 150mm 290mm, width=15cm]{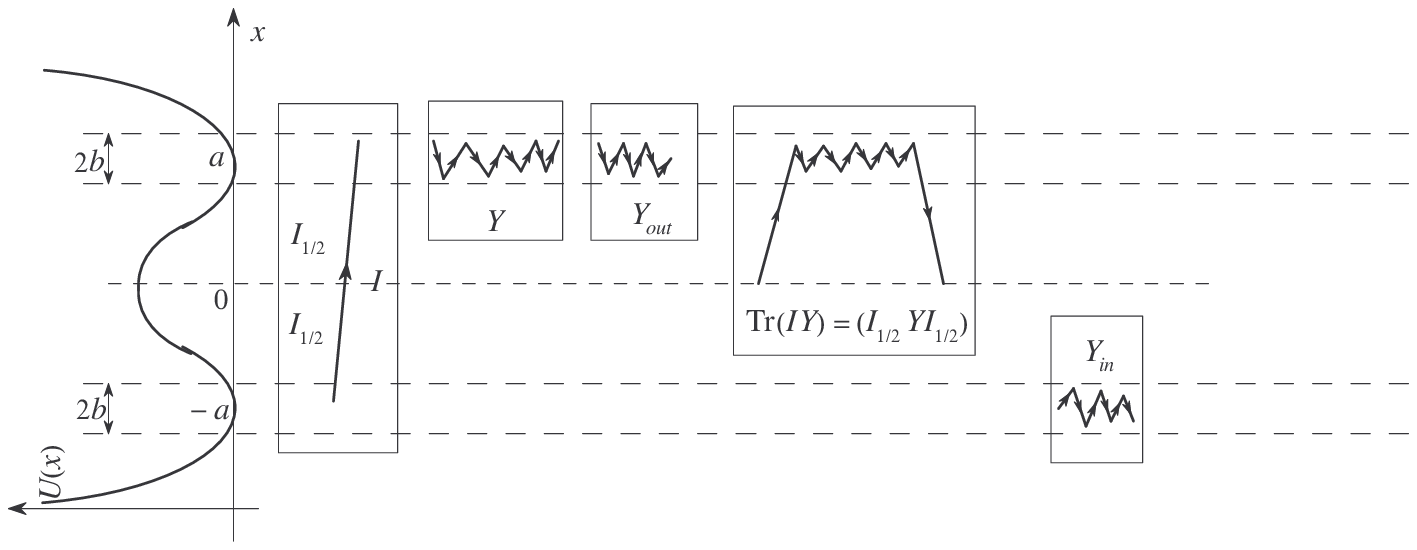}
\end{center}

With these ''blocks`` the amplitude reads (in operator form):
\begin{equation} i
K_E(-a,a)=
Y_{in}IY_{out}+Y_{in}IYIYIY_{out}+Y_{in}IYIYIYIYIY_{out}+\ldots\end{equation}

\noindent Then note that the ''instanton operator`` can be
factorized: $I(x'',x')=I_{\frac{1}{2}}(x'')I_{\frac{1}{2}}(x')$ or
in operator terms $I=I_{\frac{1}{2}}I_{\frac{1}{2}}^{\T}$. Then

\begin{equation} i K_E(-a,a)=
(Y_{in}I_{\frac{1}{2}})(I_{\frac{1}{2}}^{\T}Y_{out})+(Y_{in}I_{\frac{1}{2}})(I_{\frac{1}{2}}^{\T}YI_{\frac{1}{2}})(I_{\frac{1}{2}}^{\T}YI_{\frac{1}{2}})(I_{\frac{1}{2}}^{\T}Y_{out})+\ldots\end{equation}
\[
=(Y_{in}I_{\frac{1}{2}})\sum_{n=0}^{\infty}(\Tr{IY})^{2n}(I_{\frac{1}{2}}^{\T}Y_{out})\]

\noindent where each expression in the brackets is now just a
scalar function of $E$ (not operator). The multiplication in the
energy representation corresponds to the contraction in the time's
one:

\begin{equation} A_1(E)A_2(E)\ldots A_{n+1}(E) \leftrightarrow
\int_{\{t_{k+1}>t_k\}}\prod_{k=1}^{n}dt_kA_{n+1}(T-t_{n})A_{n}(t_{n}-t_{n-1})\ldots
A_1(t_1) \end{equation}

\noindent For $A_k(t)=\textrm{Const}\cdot e^{-\alpha t}$ this
contraction is proportional to $e^{-\alpha T}T^{n}/n!$. In our
example this integration exactly corresponds to the integration
over the centers of instantons. Described in the main text, the
semiclassical summation gives that at the most left pole
\footnote{Here $S_{inst}$ is $R$ introduced in the main text.}

\begin{equation} \Tr IY \sim
\frac{\hbar\omega/2\pi}{\frac{\hbar\omega}{2}-E}e^{-S_{inst}/\hbar}
\;\;\Rightarrow\;\;(\Tr{IY})(T)\sim \frac{\omega}{2\pi}e^{-\omega
T/2}e^{-S_{inst}/\hbar},\;T\rightarrow\infty \end{equation}

\noindent Therefore, for large $T$

\begin{equation} K_T(-a,a)=\sqrt{\frac{m\omega}{\pi\hbar}}e^{-\omega T/2}
\sum_{\parbox{1cm}{\tiny{\hbox{\hspace{0.45cm}$k=0$}\hbox{($k
\equiv 1$\hspace{-0.2cm}$\mod 2$)}}}}^{\infty}
\frac{T^k}{k!}\left(\frac{\omega}{2\pi}\right)^ke^{-kS_{inst}}=\sqrt{\frac{m\omega}{\pi\hbar}}
e^{-\omega
T/2}\sinh\left[{T\left(\frac{\omega}{2\pi}\right)e^{-S_{inst}}}\right]
\label{inst_sum}
\end{equation}

\begin{equation} \Rightarrow \Delta
E=\pm\frac{\hbar\omega}{2\pi}e^{-S_{inst}/\hbar}\end{equation}

\begin{center}
  \includegraphics[bb=10mm 245mm 185mm 290mm, width=15cm]{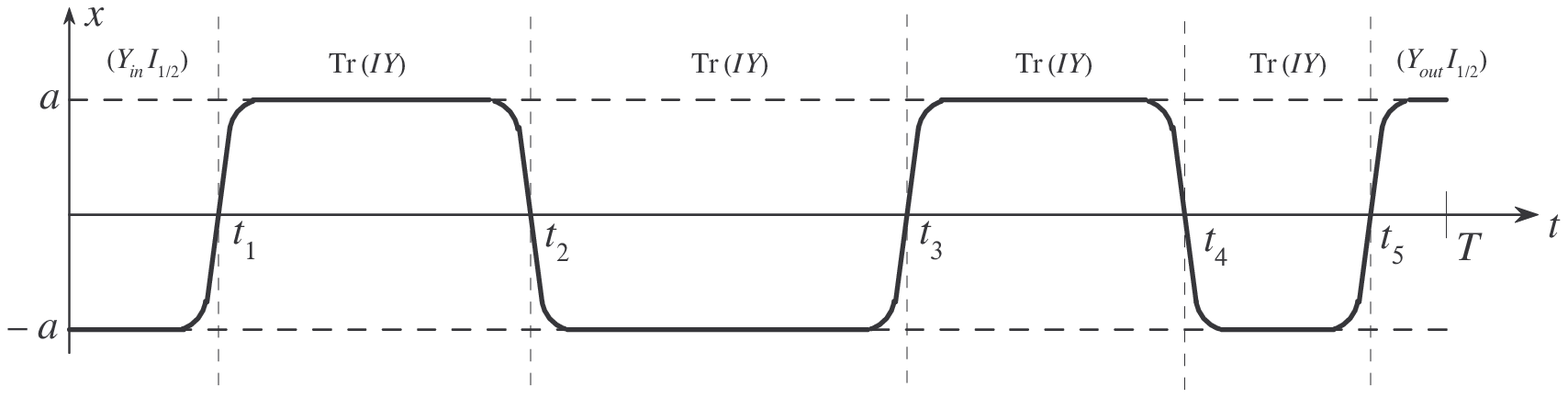}
\end{center}

\section{Perturbative test of the formula}

In this section we show how (\ref{formula_K}) agrees with the
perturbative treatment of the transition amplitude at the first
order. Namely, consider the system (we set $\hbar=1$ and $m=1$)
with $H=-\frac{1}{2}\d^2+\lambda V(x)\equiv H_0+\lambda V(x)$ and
$\lambda\rightarrow 0$. For the evolution operator $K(T)=e^{-HT}$
we have the following perturbative expansion w.r.t. $\lambda$:
\begin{equation} K(T)=K_0(T)-\lambda\int^T_0d\tau
K_0(T-\tau)V(x)K_0(\tau)+\ldots\end{equation}

\noindent with $K_0=e^{-H_0T}$. In the energy representation, it
is just a geometric series:

\begin{equation} K_0(E)=\frac{-1}{E-\lambda
V+\frac{1}{2}\d^2}=-\left(\frac{1}{E+\frac{1}{2}\d^2}+\frac{1}{E+\frac{1}{2}\d^2}\lambda
V\frac{1}{E+\frac{1}{2}\d^2}+\ldots\right) \end{equation}

\noindent We shall show that it coincides at the first order in
$\lambda$ with (\ref{oper2}). With such an accuracy we have:

\begin{equation} p^2=p_0^2\left(1-\frac{\lambda
V}{p_0^2}\right)\;\;\;\;\;\;\frac{p'}{2p}=-\frac{\lambda
V'}{2p_0^2}\;\;\;p_0\equiv \sqrt{2E} \end{equation}

\begin{equation} \frac{1}{p\pm i\d}=\frac{1}{p_0\pm i\d}+\frac{1}{p_0\pm
i\d}\frac{\lambda V}{p_0^2}\frac{1}{p_0\pm i\d} \end{equation}

\noindent In this order in $\lambda$ we need obviously only terms
in (\ref{oper1},\ref{oper2}) with less then two
reflections.\footnote{This is the general feature: to the
$\lambda^N$-order only the terms with $N$ and less reflections
contribute.}

\noindent In zero order, the agreement is trivial:
\begin{equation}
\frac{2p_0}{p_0^2+\d^2}=\frac{1}{p_0+i\d}+\frac{1}{p_0-i\d}
\end{equation}

\noindent After collecting the first-order terms, it remains to
verify that:

 \begin{equation} 4p_0^2\frac{1}{p_0^2+\d^2}V\frac{1}{p_0^2+\d^2}= \end{equation}
\[
i\left\{\frac{1}{p_0-i\d}\frac{V'}{2p_0^2}\frac{1}{p_0+i\d}-\frac{1}{p_0+i\d}\frac{V'}{2p_0^2}\frac{1}{p_0-i\d}\right\}+\]
\[+\frac{V}{2p_0^2}\frac{2p_0}{p_0^2+\d^2}+\frac{2p_0}{p_0^2+\d^2}\frac{V'}{2p_0^2}+\frac{1}{p_0+i\d}\frac{V}{p_0^2}\frac{1}{p_0+i\d}+\frac{1}{p_0-i\d}\frac{V}{p_0^2}\frac{1}{p_0-i\d}\]

\noindent We can multiply it by $(p_0^2+\d^2)$ from the left and
from the right. After algebraic simplifications, we get:

\begin{equation} 4Vp_0^2=-[\d,V']+4Vp_0^2+\{\d^2,V\}-2\d V\d \end{equation}

\noindent One can easily verify that this identity is true.

\section{Proof of the (\ref{step_alpha})} \label{tanh_proof}

In this section we shall proof that
\begin{equation} \hspace{-4mm}\int\limits_{Paths_{[a,b]}(a,a)}\hspace{-4mm}\Omega
=\sum_{n=0}^{\infty}(-1)^{n}\int_{\left\{T\geqslant
t_{2i\pm1}\geqslant t_{2i}\geqslant 0\right\} }
\prod_{k=1}^{2n+1}dt_k = \tanh T=\frac{p(b)-p(a)}{p(b)+p(a)} \label{tanh_formula_2}
\end{equation}
where we have used variables $t=\log\left[{p(x)}/{p(a)}\right]/2,\;T=\log\left[{p(b)}/{p(a)}\right]/2$. It can be analogously represented as (one should integrate over interior points from $0$ to $T$):

\begin{center}
  \includegraphics[bb=15mm 245mm 190mm 285mm, width=9cm]{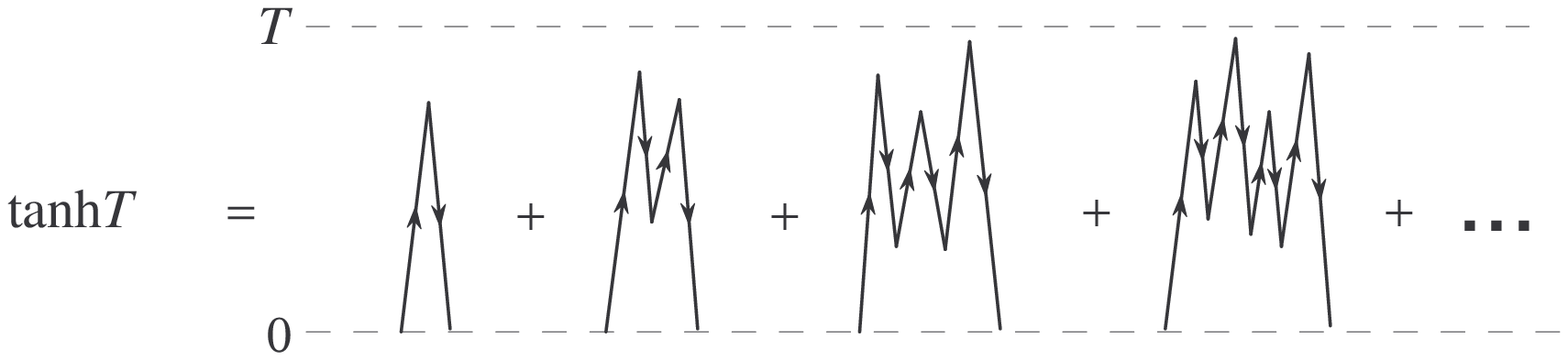}
\end{center}

 \noindent then it can be written as:
\begin{equation}\tanh{T}=G(0,0) \end{equation}
\noindent where $G(t_2,t_1)$ - kernel of operator $G$, which is the sum:
\begin{equation} G=K-K^2+K^3-K^4+\ldots=\frac{K}{1+K} \end{equation}
Operator $K$ acts on functions in the following way:

\parbox{7cm}{\begin{center}
  \includegraphics[bb=1.5cm 24.5cm 5.5cm 28.5cm, width=2.5cm]{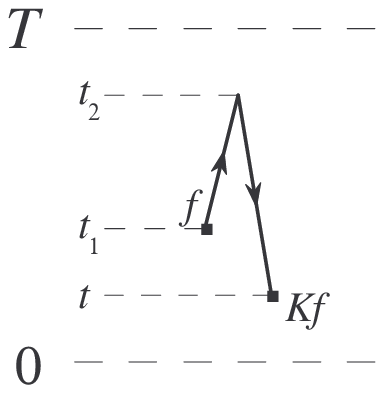}
\end{center}}
\parbox{7cm}{
\begin{equation}
(Kf)(t)=\int^T_{t}dt_2\int^{t_2}_0dt_1f(t_1)\end{equation}}

It is easy to see that $(Kf)(T)=0,\;(Kf)'(0)=0,\;(Kf)''(t)=-f(t)$.
Hence $K=-\d^{-2}$ on the space of functions $\{\phi\}$ on the segment $[0,T]$, with boundary conditions $\phi(T)=0$ and $\phi'(0)=0$. Therefore operator $G$ on this space is just 
\begin{equation} G=\frac{1}{-\d^2+1}\end{equation}
\noindent and its kernel can be represented in a standard way with the help of functions satisfying the equation ${(\d_t^2-1)\phi(t)=0}$: the first one -
$\sinh(T-t)$ - which satisfies the first boundary condition and the second one - $\cosh(t)$ - which satisfies the second boundary condition and also their Wronskian - $W=-\cosh(T)$:
\begin{equation}
G(t_2,t_1)=\frac{\cosh(t_<)\sinh(T-t_>)}{\cosh(T)}\;\;\;\;\;\;\;\;\;\;\;\;t_>=\max\{t_2,t_1\},\;t_<=\min\{t_2,t_1\}
\end{equation}
\begin{equation}
\Rightarrow G(0,0)=\tanh(T)
\end{equation}
\noindent that is what we need.

Notice that if one tries to prove (\ref{tanh_formula_2}) straightforwardly he will encounter the following combinatorial proposition: the volume of the part of the unit $(2n-1)$-dimensional cube limited by the inequalities ${\left\{1\geqslant x_{2i\pm1}\geqslant x_{2i}\geqslant 0\right\}}$, equals $\frac{2^{2n}(2^{2n}-1)}{(2n)!}B_{2n}$, where $B_k$ - Bernoulli numbers.

\section{Computations for the section \ref{schr_sect} (proof of (\ref{formula_schr_K}))} \label{schr_proof}

One can obtain the formula \begin{equation}
\int\limits_{\hbox{\tiny $Paths(x,x)$}}\hspace{-4mm}\Omega\, e^{iS}=\frac{J_R(x',x'')(1+I_L(x'))}{1-I_L(x')I_R(x')}
 \label{formula_result_K}
\end{equation}
analogously to the (\ref{Z_IR_IL}) summing 2 geometric series in the point $x'$.

\noindent Notice that if we set $x'=x''=x_0$ (using(\ref{formula_eq_J_2})) and integrate over $x_0$, we will obtain (\ref{Z_IR_IL})

It is sufficient to prove that
 \begin{equation}\frac{1}{\sqrt{p(x')p(x'')}}\frac{J_R(x',x'')(1+I_L(x'))}{1-I_L(x')I_R(x')}=-2\frac{\psi_L(x')\psi_R(x'')}{W}
\end{equation} 

to obtain (\ref{formula_schr_K}). Using (\ref{relation}), one can conclude that it is equivalent to that
\begin{equation}
\frac{i}{\sqrt{p(x')p(x'')}}J_R(x',x'')(\psi_R'(x')+ip(x')\psi_R(x'))=-2\psi_R(x'')
\end{equation}
When $x'=x''$ it is, obviously, true (substitute (\ref{formula_eq_J_2}) and (\ref{relation})). Then it remains to prove that the quantity
\begin{equation}
A(x)\equiv
\frac{J_R(x,x'')(\psi_R'(x)+ip(x)\psi_R(x))}{\sqrt{p(x)}}\end{equation}
does not depend on $x$. Let us compute its logarithmic derivative using (\ref{formula_eq_J}):
\begin{equation}
(\log
A)'=-{ip}-\frac{p'}{2p}I_R-\frac{p'}{2p}+\frac{\psi_R''+{ip}\psi_R'+{p'}\psi_R}{\psi'_R+{ip}\psi_R}
\end{equation}
After substitution of $I_R$ from (\ref{relation}) and
$\psi_R''=-p^2\psi_R$ and some algebraic simplifications one can obtain that indeed $(\log
A)'=0$.

\end{document}